\documentclass{aa}
\usepackage{graphicx}
\usepackage{natbib}
\usepackage{color}
\usepackage{float}
\usepackage{txfonts}
\usepackage{lscape}

\begin{document}

\title{Six new eccentric eclipsing systems with a third body\footnote{Table XXX is only available in electronic form at the CDS via anonymous ftp to cdsarc.u-strasbg.fr (130.79.128.5) or via http://cdsweb.u-strasbg.fr/}}

\author{Zasche,~P.~\inst{1},
        Henzl,~Z.~\inst{2,3},
        Wolf,~M.~\inst{1}}

\offprints{Petr Zasche, \email{zasche@sirrah.troja.mff.cuni.cz}}

 \institute{
  $^{1}$ Astronomical Institute, Charles University, Faculty of Mathematics and Physics, V~Hole\v{s}ovi\v{c}k\'ach 2, CZ-180~00, Praha 8, Czech Republic\\
  $^{2}$ Variable Star and Exoplanet Section, Czech Astronomical Society, Fri\v{c}ova 298, CZ-251 65 Ond\v{r}ejov, Czech Republic\\
  $^{3}$ Hv\v{e}zd\'arna Jaroslava Trnky ve Slan\'em, Nosa\v{c}ick\'a 1713, CZ-274 01 Slan\'y 1, Czech Republic\\
 }

\titlerunning{Six new eccentric triple systems}
\authorrunning{Zasche \& Henzl}

  \date{\today}

\abstract{We present the discovery of six new triple stellar system candidates composed of an inner
eccentric-orbit eclipsing binary with an apsidal motion. These stars were studied using new, precise
TESS light curves and a long-term collection of older photometric ground-based data. These data were
used for the monitoring of ETVs (eclipse timing variations) and to detect the slow apsidal movements
along with additional periodic signals. The systems analysed were ASASSN-V~J012214.37+643943.3 (orbital
period 2.01156~d, eccentricity 0.15, third body with 3.3~yr period); ASASSN-V~J052227.78+345257.6
(2.42673~d, 0.35, 3.2~yr); ASASSN-V~J203158.98+410731.4 (2.53109~d, 0.20, 2.7~yr);
ASASSN-V~J230945.10+605349.3 (2.08957~d, 0.18, 2.3~yr); ASASSN-V~J231028.27+590841.8 (2.41767~d, 0.43,
4.9~yr); and NSV 14698 (3.30047~d, 0.147, 0.5~yr). In the system ASASSN-V~J230945.10+605349.3, we
detected a second eclipsing pair (per 2.99252~d) and found adequate ETV for the pair B, proving its 2+2
bound quadruple nature. All of these detected systems deserve special attention from long-term studies
for their three-body dynamics since their outer orbital periods are not too long and because some
dynamical effects should be detectable during the next decades. The system NSV~14698 especially seems
to be the most interesting from the dynamical point of view due to it having the shortest outer period
of the systems we studied, its fast apsidal motion, and its possible orbital changes during the whole
20th century. }


\maketitle

\section{Introduction} \label{intro}

The role of eclipsing binaries in current astrophysics is undisputed since these objects still provide
us with the most precise method for deriving stellar masses and radii. Various studies have also
discussed their advantages for calibrating the stellar evolutionary models (see e.g.
\citealt{2012ocpd.conf...51S}) as well as their use for studies of multiples among the stellar
population \citep{2022Galax..10....9B, 2018ApJS..235....6T}. In particular, eccentric-orbit eclipsing
binaries where the slow apsidal motion of both eclipses is detectable play a significant role in the
testing of stellar models and proving the models of general-relativity apsidal motion
\citep{2019A&A...628A..29C}. This is even more evident thanks to new, super-precise TESS data for this
type of system \citep{2021A&A...654A..17C}.

The whole situation can be quite more complicated when a system contains a third body. In some cases,
besides the observed apsidal motion, another periodic modulation can also be detected, indicating the
presence of a distant third component in a system \citep{2007MNRAS.379..370B}. In their study,
\citeauthor{2007MNRAS.379..370B} only dealt with long, periodic outer bodies (typically years to
decades), and the influence of these third bodies on the inner pair of a system was almost negligible.
A much more interesting situation occurs when the outer body orbits at a closer distance to the
eclipsing pair or when the ratio of outer to inner periods $p_3/P_s$ is adequately low. The complex
dynamics of such systems should show rather interesting eclipse timing variation (ETV) diagrams (as
shown, e.g., in the study of Kepler eccentric eclipsing binaries with close third bodies in
\citealt{2015MNRAS.448..946B}). On longer timescales, one can expect triple dynamics and the impact of
the Kozai-Lidov cycles (\citealt{1962AJ.....67..591K} \& \citealt{1962P&SS....9..719L}), for example,
on the dynamical evolution of the orbits, such as its precession and inclination changes. We believe
that only when the number of such triple systems increases can their origin and formation mechanism, as
well as also their subsequent orbital evolution, be studied. Much more robust samples of these systems
can bring some clue to such open questions as whether the systems are products of disc fragmentation or
some N-body dynamics.

It is quite interesting how limited the sample of eccentric apsidal motion systems with a third body
was just a few years ago. The comprehensive catalogue compiling all known galactic eclipsing systems
with eccentric orbits published by \cite{2018ApJS..235...41K} shows in all 139 systems with only
apsidal motion and 31 systems that exhibit both apsidal motion and the light-travel time effect
(hereafter, LTTE; see e.g. \citealt{1952ApJ...116..211I}, or \citealt{1990BAICz..41..231M}; for more
tight systems with shorter outer periods see e.g. \citealt{2015MNRAS.448..946B}). Also quite remarkable
is the fact that outside of our Galaxy, there were announced even more detections of candidate triples
using the same methods of ETV analysis and using mainly OGLE data \citep{1997AcA....47..319U} spanning
several years. The situation has rapidly changed due to recent missions such as Kepler
\citep{2010Sci...327..977B} and TESS \citep{2015JATIS...1a4003R}. Dozens of new triples have been
discovered, mainly on very short outer periods, causing very strong dynamical ETV effects due to
non-negligible third-body dynamics. (See the summary and discussion sections for such systems.)
Therefore, any new eclipsing system showing an eccentric orbit with apsidal motion together with a
third-body indication via LTTE would be very welcome. And this was the main motivation for our study.

\section{Method}

For the analysis of the selected systems, we mainly used the super-precise data from the TESS
satellite. The light curves (hereafter, LCs) for the individual stars were extracted from the raw data
using the programme {\tt{lightkurve}} \citep{2018ascl.soft12013L}. Depending on the brightness of the
individual star, we also used the appropriate aperture (number of TESS pixels) to extract the
photometry. Some basic information about the studied systems is summarised below in Table
\ref{systemsInfo}.

We only selected systems where the apsidal advance and additional variation are clearly visible
(even only in the TESS photometry). Then, confirmation of the additional ETV was done using older,
independent ground-based data to definitively prove the triple-star hypothesis. For this
confirmation, several different sources of data were used. These sources were mainly the ASAS-SN
survey \citep{2014ApJ...788...48S,2017PASP..129j4502K}, ZTF \citep{2019PASP..131a8003M}, SuperWASP
\citep{2006PASP..118.1407P}, Atlas \citep{2018AJ....156..241H}, NSVS \citep{2004AJ....127.2436W},
and KELT \citep{2007PASP..119..923P,2018AJ....155...39O}.

Our method was the following. From the precise TESS data, we derived the light curve fit using the {\sc
PHOEBE} software \citep{2005ApJ...628..426P}. From the less precise ground-based data, we only derived
the individual times of the eclipses in order to trace the slow apsidal motion of the system and to
detect the ETVs. For the computation of the eclipse times, we used our method of AFP as described in
our paper \cite{2014A&A...572A..71Z}, which involves using the LC template to derive the times of
eclipses of both the primary and secondary. All of the derived times of eclipses will be made available
as online-only material via CDS.

We then analysed the whole set of eclipse times in a standard way. This means we tried to minimise the
sum of square residuals from the fit. Such a fit incorporates the ephemerides of the inner binary (i.e.
$HJD_0$, $P_s$), the apsidal motion of the orbit ($\omega_0$, $e$, $\dot\omega$), and the LTTE
parameters ($p_3$, $A$, $e_3$, $\omega_3$, $T_0$). All ten parameters had to be fitted simultaneously.
A question about the eccentricity value (and $\omega$ as well as $\dot\omega$) naturally arose because
all of these values have to be used for both the light curve fitting and the ETV fitting. We used the
following procedure. At first, the eccentricity value was taken from the preliminary ETV fit. This
value was then taken as input for the LC fitting, and a new eccentricity value from the LC was then
taken as input for the new ETV fit. The $e$ values derived from both methods were then compared, and
they usually agreed well with each other. However, if there was some disagreement among them, then some
reasonable average value between the two was taken, and with that value, the whole analysis was redone.

\section{Results for individual systems}

In this section, we only briefly mention the individual systems that we analysed. In general, the
methods used to analyse each system are very similar to each other.

\subsection{ASASSN-V J012214.37+643943.3}

The very first system, named ASASSN-V~J012214.37+643943.3, was discovered as an eclipsing binary a few
years ago based on the ASAS-SN data \citep{2019MNRAS.486.1907J}. Its orbital period is about two days,
and the primary eclipse appears slightly deeper than the secondary one. It has an eccentric orbit that
is even visible with the naked eye, thanks to its shift of the secondary eclipse away from the 0.5
phase. However, the star has not been studied before, and thus no light curve analysis has been
performed; nobody noticed its apsidal motion. Unfortunately, spectrum for the star cannot be found in
available databases.

From the available TESS sectors of data (there are currently five sectors), we chose sector 58 for the
LC analysis due to its shortest exposing time (i.e. the best data coverage and largest number of data
points). Using the  {\sc PHOEBE} programme, we carried out the analysis. Results of the LC fitting are
given in Table \ref{LCparam}. Quite surprisingly, the results showed that the light contribution is
very similar for all three components in the system. However, this does not necessarily mean that the
third body in the system is about the same mass as the inner components of the eclipsing pair. Due to
large TESS pixels, the light contribution from the close-by sources cannot be ignored and should be
taken into account (a few close <15" sources exist, but with much lower luminosity).

Concerning its period changes, a slow apsidal motion of the inner eccentric orbit was detected. The
movement is rather slow, with an apsidal period of about 203 years. In addition, a short-term variation
with a periodicity of 3.3 yr was also derived. This evident variation on the five sectors of TESS data
is now fairly well supported based on the older ground-based observations. Our five new observations of
eclipses also provide support for this short-term variation (see Figure \ref{figures} with our new data
in red). The orbit seems to be circular, and from the resulting parameters of the LTTE fit, one can
make at least some rough conclusions. For example, when assuming a co-planar configuration, the third
body should have the same mass as the eclipsing components (i.e. its 1/3 luminosity should be in good
agreement). From its parallax from GAIA, the predicted angular separation of the components can also be
estimated, which we found to be about 2~mas and should be taken into account for a prospective
interferometric detection in the future.

 \subsection{ASASSN-V J052227.78+345257.6}

Another ASAS-SN discovery was the star ASASSN-V J052227.78+345257.6, which has about 2.4-day period and
was also not studied before. Just as with the previous star, we collected all the data available for
our analysis. Thanks to the brightness of the star, there are much more older data, which helped us a
lot in tracing the slow apsidal motion.

We used the TESS sector 59 for the modelling of the LC. The final parameters are given in Table
\ref{LCparam}. In the table, one can see that the level of the third light is significantly lower (it
contributes only about 10\% to the total light), and both components are much more detached than in the
previous system. Quite surprisingly, we report an eccentricity of about 0.35, which is one of the
highest values for the stars with periods below 2.5~days.

\begin{figure*}
   \centering
   \includegraphics[width=0.98\textwidth]{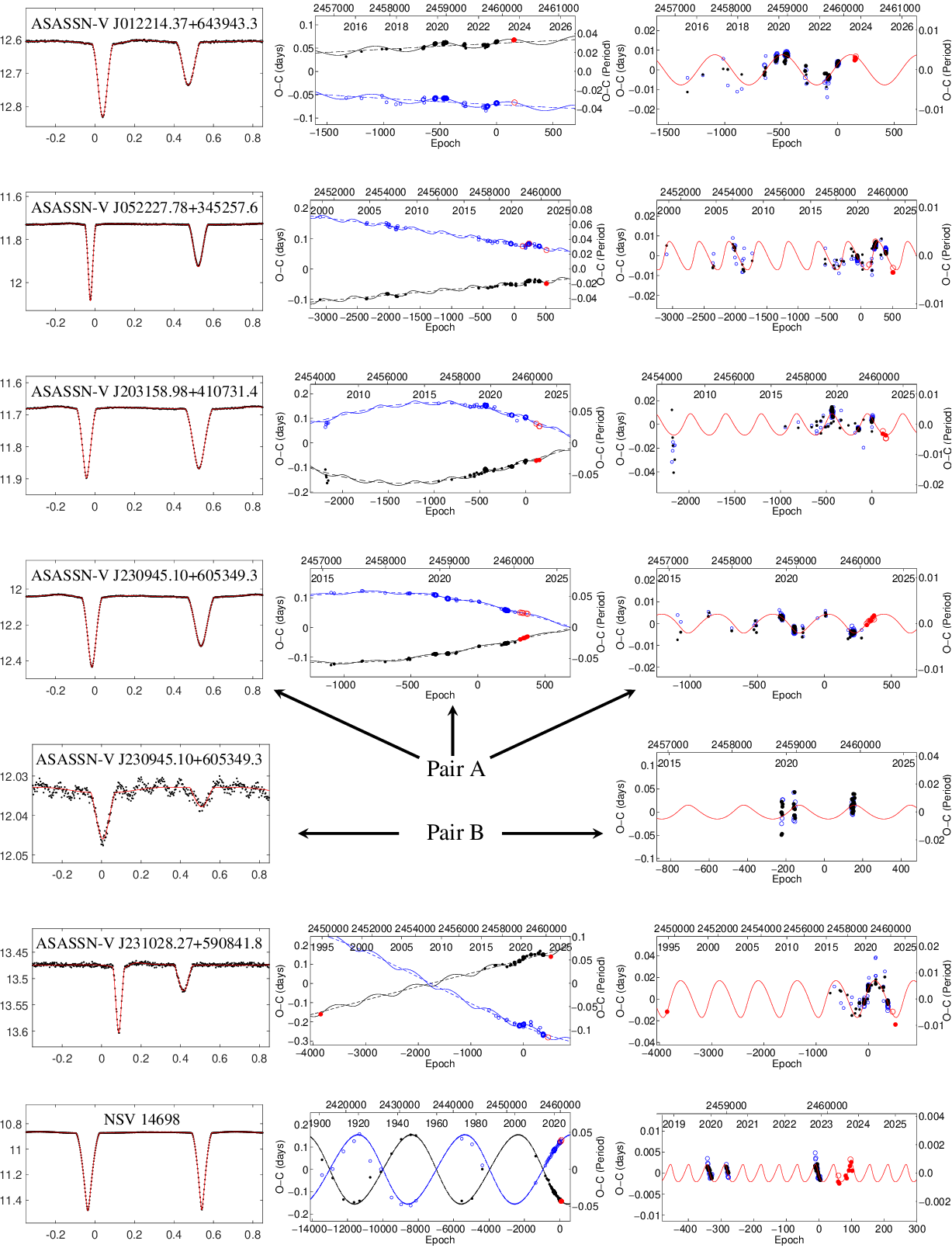}

   \caption{Analysed systems and their fits. Each row of figures represents one system. The figures in the
   left column are the light curve fits from TESS data, while in the right column are the diagrams showing
   the period changes. Middle plots show the complete fit (apsidal motion plus LTTE), right-hand plots are only LTTE fits.
   Black dots indicate primary eclipses, blue dots are the secondary ones, while the red data show our
   newly dedicated observations. The bigger the symbol, the higher the weight and precision with which it
   was derived.}
   \label{figures}
  \end{figure*}

We were able to derive the times of eclipses from the data spanning back more than 20~years. This is
mainly due to the star's rather deep eclipses. Five new eclipses were also observed with our telescopes
especially for this study. Analysing all of these available data points resulted in the the picture
shown in Table \ref{OCparam} and Figure \ref{figures}). The apsidal motion has a period of about 450
years, while the shorter ETV variation visible on both the primary and secondary eclipses has a
periodicity of about 3.2~years only. With the putative third body having such a short orbit, one can
also ask whether some dynamical shorter-period perturbations would be detectable there on the
human-life timescale. When computing the rough estimate of such effects (see e.g.
\citealt{2016MNRAS.455.4136B}), the ratio of $p_3^2/P_s$ gave us an order of magnitude estimation of
the period of such a modulation. It resulted in about 1500 years, which is short enough to be
considered. For example, the most pronounced effect should be the precession of the orbital planes
(i.e. a possible change of the eclipse depth of the inner eclipsing pair may be detectable during the
next decades).

%

 \begin{table*}
   \caption{List of new eccentric triple candidates.}  \label{systemsInfo}
   \scalebox{0.92}{
   \begin{tabular}{c c c c c c c }\\[-2mm]
 \hline \hline
     RA      &     DE      & VSX Target name              & TESS number    & T$_{mag}$&  GAIA plx     &  Spectral/temperature  \\[0.4mm]
  [J2000.0]  &  [J2000.0]  &                              &                &          &   [mas]       &     information$^{*}$       \\[0.4mm]
  \hline
  01 22 14.4 & +64 39 43.3 & \object{ASASSN-V J012214.37+643943.3} & TIC 444635304 &  12.603   & 0.491 (0.012) &  GAIA: 9327.4 K  \\
  05 22 27.8 & +34 52 57.6 & \object{ASASSN-V J052227.78+345257.6} & TIC 2775707   &  11.729   & 1.181 (0.030) &  GAIA: 9372.3 K  \\
  20 31 59.0 & +41 07 31.4 & \object{ASASSN-V J203158.98+410731.4} & TIC 17033439  &  11.678   & 0.561 (0.017) &  sp B1V$^\sharp$ \\
  23 09 45.1 & +60 53 49.3 & \object{ASASSN-V J230945.10+605349.3} & TIC 268527079 &  12.033   & 0.358 (0.014) &  GAIA: 5856.5 K  \\  
  23 10 28.3 & +59 08 41.8 & \object{ASASSN-V J231028.27+590841.8} & TIC 315513397 &  13.474   & 0.200 (0.020) &  GAIA: B$_p$-R$_p$= 1.252 mag \\
  23 46 08.6 & +62 03 01.2 & \object{NSV 14698}                    & TIC 272624786 &  10.864   & 0.302 (0.014) &  sp B2 $^\ddag$     \\
  \hline
 \end{tabular}}\\[1mm]
  {\scriptsize Notes: $^{*}$ - GAIA taken from release DR3 \citep{2016A&A...595A...1G,2023A&A...674A...1G},  $^\sharp$ - taken from \cite{2018A&A...612A..50B}, $^\ddag$ \cite{1953IzKry..10..104B} }
 \end{table*}

\begin{table*}
\caption{Light curve parameters for the analysed systems.} \label{LCparam}
\scalebox{0.82}{
\begin{tabular}{lccccccccc}
  \hline\hline\noalign{\smallskip}
        System                &  $T_1$ [K]   &  $T_2$ [K]  &  $i$ [deg]   &  $\Omega_1$   &  $\Omega_2$   &  $L_1$ [\%] &  $L_2$  [\%] & $L_3$ [\%] \\ 
  \hline\noalign{\smallskip}
 ASASSN-V J012214.37+643943.3 & 9327 (fixed) &  8715 (157) & 79.13 (0.54) & 6.333 (0.038) & 6.308 (0.048) & 37.1 (0.5)  &  31.3 (1.7)  & 31.6 (0.9) \\ 
 ASASSN-V J052227.78+345257.6 & 9372 (fixed) & 10149 (99)  & 83.95 (0.49) & 8.722 (0.029) &10.283 (0.056) & 49.9 (0.4)  &  40.0 (1.9)  & 10.1 (3.1) \\ 
 ASASSN-V J203158.98+410731.4 &25000 (fixed) & 25246 (274) & 84.26 (0.84) & 6.741 (0.051) & 6.400 (0.035) & 22.2 (0.8)  &  27.0 (1.2)  & 50.8 (6.4) \\ 
 ASASSN-V J230945.10+605349.3 & 5857 (fixed) &  5747 (116) & 81.79 (0.66) & 6.249 (0.039) & 6.392 (0.041) & 47.4 (0.9)  &  43.8 (1.4)  &  8.8 (3.2) \\ 
 ASASSN-V J231028.27+590841.8 & 4853 (fixed) &  5284 (107) & 81.25 (0.95) & 9.582 (0.088) & 8.155 (0.063) & 14.6 (0.7)  &  25.3 (3.7)  & 60.1 (8.0) \\ 
 NSV 14698                    &20000 (fixed) & 19792 (126) & 88.11 (0.35) & 6.834 (0.046) & 6.970 (0.059) & 49.5 (0.5)  &  45.7 (0.8)  &  4.8 (1.1) \\ 
 \noalign{\smallskip}\hline
\end{tabular}}
\end{table*}

\begin{table*}
\caption{Parameters from the ETV analysis.} \label{OCparam}
 \scalebox{0.7}{
\begin{tabular}{l|cc|ccc|ccccc}
\hline\hline 
         System               & $HJD_0 - 2400000$ &    $P_s$       &   $e$     & $\dot{\omega}$  & $\omega_0$  &  $p_3$    &    $A$     &    $e_3$    &  $\omega_3$ &   $T_0 - 2400000$    \\
                              &      [HJD]        &     [days]     &           & [deg$/\rm{yr}$] &   [deg]     &   [yr]    &   [day]    &             &             &    [HJD]   \\
 \hline 
 ASASSN-V J012214.37+643943.3 &  59894.5136 (32)  & 2.0115580 (51) & 0.152 (11)&   1.77 (0.23)   & 132.5 (1.3) & 3.30 (0.80)& 0.008 (3) & 0.00 (0.01) & 136.9 (52.8)& 61669 (320) \\
 ASASSN-V J052227.78+345257.6 &  58966.0766 (28)  & 2.4267283 (39) & 0.353 (18)&   0.79 (0.20)   &  75.9 (2.0) & 3.15 (0.35)& 0.007 (2) & 0.40 (0.05) & 12.6 (2.0)  & 61744 (415) \\
 ASASSN-V J203158.98+410731.4 &  59811.4145 (21)  & 2.5310896 (74) & 0.202 (17)&   6.94 (0.71)   &  57.1 (3.1) & 2.68 (0.76)& 0.009 (3) & 0.28 (0.10) & 67.5 (10.3) & 58660 (293) \\
 ASASSN-V J230945.10+605349.3 &  59439.2181 (39)  & 2.0895677 (95) & 0.178 (23)&   9.59 (0.96)   &  50.3 (2.5) & 2.34 (0.89)& 0.005 (2) & 0.27 (0.11) & 357.1 (15.4)& 54973 (317) \\
 ASASSN-V J231028.27+590841.8 &  58965.4267 (85)  & 2.4176687 (101)& 0.434 (22)&   2.69 (1.32)   & 121.5 (5.2) & 4.92 (0.94)& 0.017 (5) & 0.22 (0.10) & 313.3 (14.7)& 58619 (433) \\
 NSV 14698                    &  59895.9645 (86)  & 3.3004715 (16) & 0.147 (7) &   6.60 (0.10)   & 326.4 (1.2) & 0.50 (0.02)& 0.002 (1) & 0.43 (0.03) & 128.9 (3.1) & 58773 (19)  \\
 \noalign{\smallskip}\hline
\end{tabular}}
\end{table*}

   \begin{figure}
   \centering
   \includegraphics[width=0.45\textwidth]{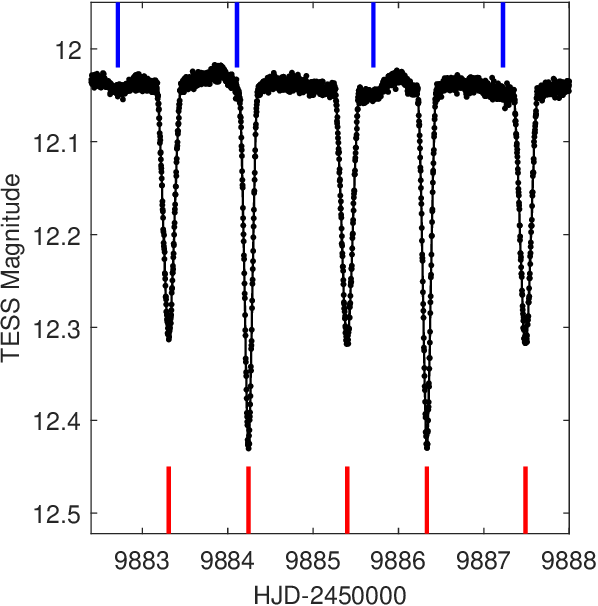}
   \caption{Photometry of ASASSN-V J230945.10+605349.3 as obtained during sector 58 of TESS data. Deeper eclipses are from pair A (denoted with red abscissae), while shallower eclipses from pair B (blue abscissae) are only barely visible.}
   \label{figASAS230945}
  \end{figure}

 \subsection{ASASSN-V J203158.98+410731.4}

Next in our sample of stars is ASASSN-V J203158.98+410731.4, also known as RLP~961. It is probably the
most studied system within our group of stars, partly due to its brightness. It was classified as a B1V
star and a member of Cygnus OB2 association \citep{2018A&A...612A..50B}.

Our LC analysis was performed on the TESS light curve from sector 55. The results of the fitting are
given in Table \ref{LCparam}, and the fit is plotted in Figure \ref{figures}. As one can see, we
interchanged the role of the two components (i.e. as the secondary component is more luminous, it has a
larger radius and higher mass). This comes from the fact that we tried to place the deeper eclipse
closer to the zero phase in the phased light curve. However, quite surprisingly, the additional third
light value we found is in more than 50\% of the total light.

We collected all available data for the period analysis. Apart from what came from TESS, the data
included the observations from surveys ZTF, ASAS-SN, and SuperWASP. In addition, we also observed
the star with our telescopes and derived five new times of eclipses, which follow the prediction
quite well, as shown in Figure \ref{figures}. The apsidal motion resulted in a rather short
apsidal period of about 52 years. Additional variation has a period of 2.7 years only. Due to its
short outer orbit, the ratio $p_3^2/P_s$ resulted in about 1030 years, which is short enough to be
detected in case of non-coplanar orbits (e.g. via an orbital precession), while the predicted
angular separation of about 2~mas is too small to resolve an 11-mag star using current
instruments.

 \subsection{ASASSN-V J230945.10+605349.3}

The stellar system ASASSN-V J230945.10+605349.3 is probably the most interesting in our sample of
stars. The never-before-studied star shows a detached-like LC with rather deep eclipses (the primary
eclipse 0.4 mag deep), while its period is about 2.1 days. No spectral analysis was performed; however,
even the low resolution GAIA spectra shows some weak indication of a double-peaked profile.

For the LC analysis, we used the TESS data from sector 58, which evidently shows a deeper primary but a
wider secondary eclipse (see Fig.\ref{figures}). Both eclipsing components are quite similar to each
other (see Table \ref{LCparam}), while the value of the third light only indicated a small fraction of
the third light below 10\%. The most surprising outcome in the analysis was the discovery of another
variability on the TESS light curve (see Figure \ref{figASAS230945}). Very shallow additional eclipses
were detected there (on all four sectors of data) having a period of 2.9925 days and a depth only about
1/30 of the of depth of the primary eclipses of pair A. Moreover, even shallower pulsational-like
variability was visible on the B light curve, having a period of about 0.3 days. Very preliminary LC
analysis of pair B indicated that its light contribution is really very small compared to the pair A
(i.e. about 91\% of the light comes from A and 9\% from pair B). However, with only this statement, one
cannot be sure that the two A and B binaries really constitute a bound quadruple system.

A quite straightforward way to prove the gravitational coupling is an ETV analysis of both pairs. We
used a method similar to the one in our recent publications detecting several new doubly eclipsing
systems through ETV analysis of both pairs and ground-based photometric data from various surveys (see
e.g. \citealt{2023A&A...675A.113Z}). The pair A shows an evident and quite rapid apsidal motion with a
period of about 37 years, and the eccentricity was derived to be 0.177. Figure \ref{figures} shows our
final fit with the combination of apsidal motion and the LTTE hypothesis. However, for pair B, any such
analysis was very difficult to reproduce. Pair B has only very shallow eclipses, and these cannot be
detected in the ground-based data. Due to this, we only used the TESS data available from four sectors
and plotted the times of eclipses of pair B to the ETV diagram of B below that of A with the same range
of x-axis in order for them to be comparable. As one can see, there is evident variation of pair B's
eclipses in the anti-phase with respect to pair A's eclipses, clearly showing that the star is a bound
quadruple 2+2 stellar system. If one has adequately good data with times of eclipses for both the A and
B pairs, one is in principle able to derive their ETV amplitudes and therefore also the mass ratio
$M_A/M_B$ of both pairs. However, the ETV here for pair B is so poorly constrained that the amplitude
of variation of B could hardly be estimated. However, it does seem to be higher than for A, which is in
agreement with the luminosity ratio of both A and B pairs.

 \subsection{ASASSN-V J231028.27+590841.8}

The eclipsing system ASASSN-V J231028.27+590841.8 is the never-before-studied star. Its period is about
2.4 days, and it shows two rather incomparable eclipses. The primary one is narrow and about 0.12 mag
deep, while the secondary eclipse is obviously longer in duration and less than half as deep as the
primary one.

Due to a lack of temperature information from the GAIA DR3 (it is the faintest source in our sample of
stars), we found it quite problematic to at least roughly estimate the temperature for the primary
component for the LC modelling. Based on the photometric index GAIA DR3: $(B_p-R_p)$= 1.252~mag, we
drew the conclusion (when ignoring the interstellar extinction) that the source is probably some dwarf
star of the early K type. However, the index $(J-H)$ from 2MASS \citep{2006AJ....131.1163S} is
0.275~mag, indicating a rather earlier type of early G star. On the other hand, the $(B-V)$ index from
APASS \citep{2015AAS...22533616H} is 0.869~mag, showing an early K star. This would be in good
agreement with the previous estimate of $T_{eff}$ as 4853~K given by GAIA DR2
\citep{2018A&A...616A...1G}.

Sector 57 was used for the LC modelling of the eclipsing pair. The results are given in Table
\ref{LCparam}, where one can see that the secondary component is also the more dominant one, both in
temperature and radius (hence also in luminosity). A remarkably large fraction of the third light
resulted from our LC modelling, suggesting a more massive third component in the system (there is no
obvious close-by companion in the vicinity of the target that would contaminate the large TESS pixel).
What is even more surprising here is the high value of eccentricity. With its 2.4-day period and
$e=0.434$, it is one of the record-breaking systems due to it having such a short orbital period.

Collecting the available data from various sources (ASAS-SN, ZTF, Atlas), we carried out a period
analysis for the apsidal motion analysis. Its period of about 136 years is still only poorly
constrained with data. However, the short-period, additional variation in the ETV diagram (see Figure
\ref{figures}) is now clearly visible, especially thanks to our dedicated observations. Luckily, one
older observation was found (due to the proximity of the star to a known eccentric system
\object{PV~Cas}), which was obtained in Ond\v{r}ejov observatory in 1994. Our data baseline was spread
significantly by this one very useful data point. With two new observations, we arrived at a final
picture of the whole system, presented in Figure \ref{figures}. The period of the ETV is about 4.9
years, longest in our sample.

\subsection{NSV 14698}

The last system we studied was NSV~14698. It is the brightest star in our sample and has the longest
orbital period, about 3.3 days. It was discovered as a variable by \cite{1958MmSAI..29..465R}, and its
spectral type was derived as B2 by \cite{1953IzKry..10..104B}. However, no detailed analysis of the
system has been published until now.

We took the TESS sector 58 for the LC analysis, resulting in the parameters given in Table
\ref{LCparam} and the final plot in Figure \ref{figures}. As one can see, the two eclipsing components
are rather similar to each other. This system exhibits the deepest eclipses among our sample of stars,
which are caused by its large inclination angle being rather close to 90$^\circ$. The third light value
resulted in only a small fraction of the total light.

We collected all available data for the ETV analysis, which spans now more than 120 years. Due in
particular to the system's deep eclipses, it was possible to derive many useful observations from
photographic plate archives, such as  DASCH \citep{2013PASP..125..857T}. (For the complete analysis of
all these data, see Figure \ref{figures}). As one can see from Figure \ref{figures}, several periods of
the apsidal motion are covered (its period is about 54 years). However, from the most recent data
points with higher precision, another short-period and smaller-amplitude variation of LTTE can also be
seen. We plotted only the more precise observations from TESS and our dedicated observations in the
right-hand plot in Figure \ref{figures}. This small variation cannot be seen in older data due to its
small amplitude (about 3 minutes only, which is the smallest LTTE amplitude in our sample) and due to
its quite short orbital period (about half a year).

Such a short periodic variation is remarkable since most of the similar eccentric triple systems show
much longer periods. Therefore, it would be expected to also take into account the dynamical effects of
the third body in addition to the classical geometrical LTTE (similar to what was done in, e.g.,
\cite{2015MNRAS.448..946B}). However, due to its rather complicated fitting, we decided to use only the
simplified LTTE approach while considering the limitations of such an approach. Our model therefore
cannot provide realistic parameters. Rather, it is to be taken as proof that this star is really the
most dynamically interacting in our sample and to possibly attract special attention to this system,
both for observational as well as for modelling efforts. We computed that due to the low ratio of
periods $p_3^2/P_s$ (about only 28 years), the ratio of amplitudes for both ETVs (LTTE and dynamical)
result in about $A_{LTTE}/A_{phys} \approx 0.027$. This shows that the dynamical effect absolutely
dominates here and that a more sophisticated approach for modelling is needed. We should also speculate
about some possible inclination changes due to the third-body effects in the system. However, due to it
having detectable eclipses during the whole 20th century, one can suspect that the system is probably
close to co-planar and that these dynamical effects are small enough. However, what do not fit very
well are the oldest data points from the first half of the 20th century. Having a very well-defined
apsidal motion with the recent data from the last 20 years, one would expect that the older
observations would also be adequately well described with such an apsidal motion. However, one can see
there is some small deviation of the older points. The most probable explanation for this discrepancy
is the dynamical influence of the distant body and our inadequate fit.

\section{Conclusions}

\begin{figure}
 \centering
 \begin{picture}(380,125)
 \put(0,2){
  \includegraphics[width=0.23\textwidth]{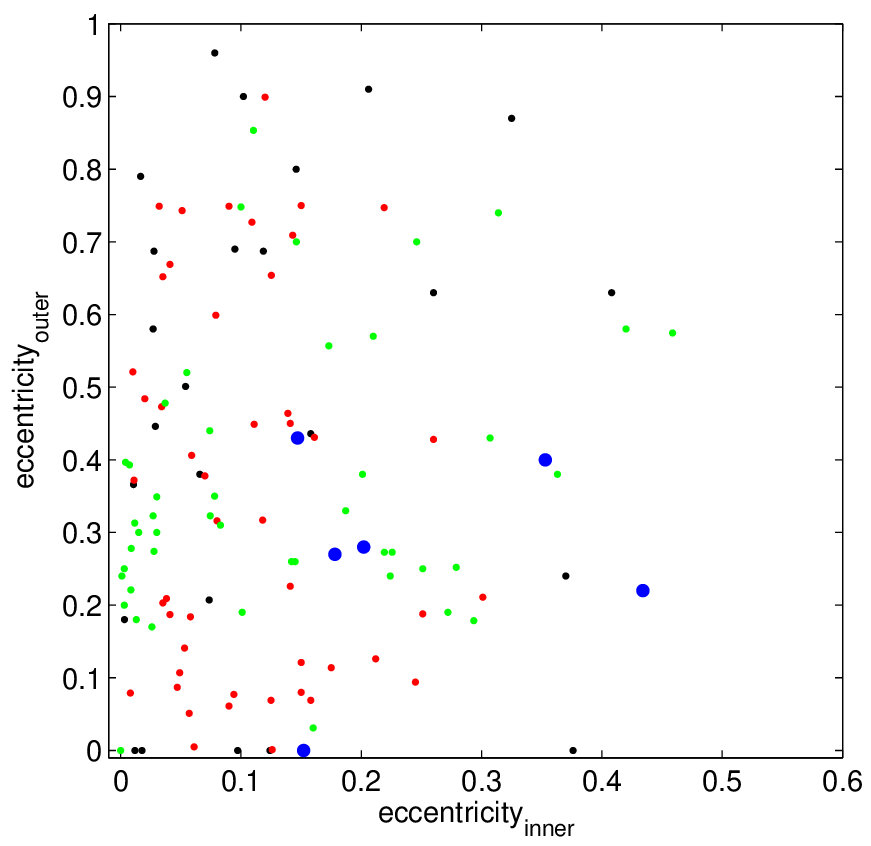}}
  \put(130,0){
  \includegraphics[width=0.23\textwidth]{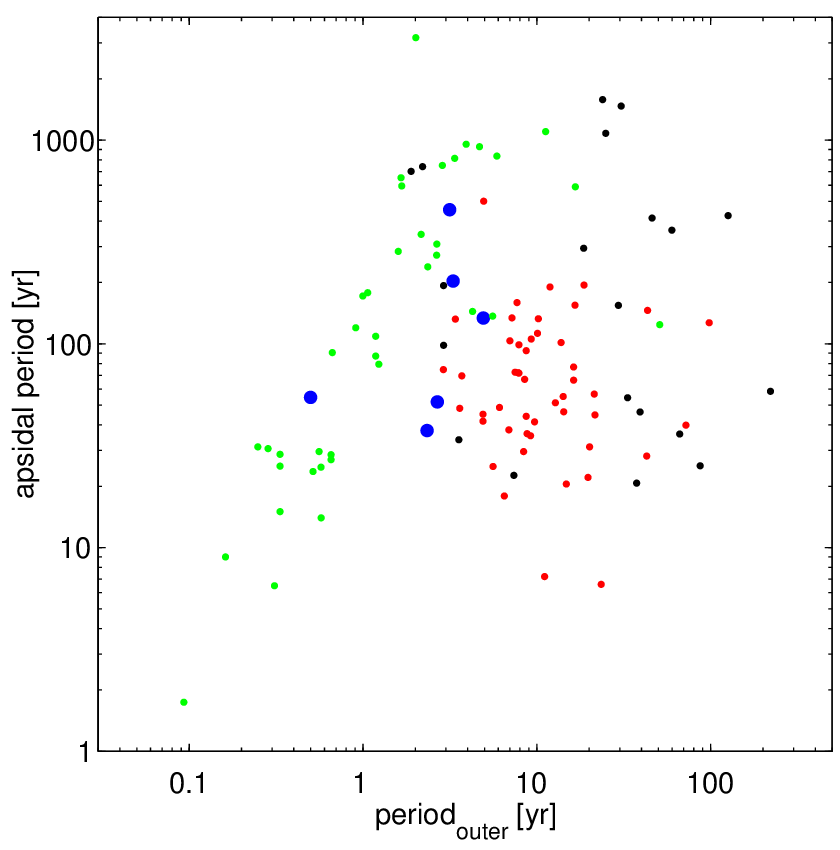}}
  \end{picture}
  \caption{Distribution of orbital parameters of currently known and published eccentric apsidal motion triples.
  The black dots denote well-known classical apsidal systems in our Galaxy; red dots are the stars in LMC/SMC;
  green dots denote new, recently published systems based on the new TESS and Kepler data; and the blue dots
  are the six new systems we present.}
  \label{Fig_ecc_triples}
 \end{figure}

We have carried out the very first analysis of six unstudied eclipsing eccentric systems with
putative third bodies. These systems are of great importance to deepening our knowledge about the
multiple systems, especially the dynamics of stellar triples. For this purpose, we plotted all
known eclipsing systems showing some apsidal motion together with additional ETV variation
interpreted as a third-body influence (see Fig.\ref{Fig_ecc_triples}). Apart from the classical,
well-known, and often studied systems (typically with GCVS names), we also plotted recent
discoveries based on the new, precise TESS and Kepler data, which usually covers a much shorter
time interval. Therefore, these systems typically show short periods for the third bodies, and the
third components strongly and dynamically interact with the inner eclipsing pairs (see the green
dots in Figure \ref{Fig_ecc_triples}.  One can see that there is no obvious correlation between
the inner and outer eccentricities. On the other hand, the system NSV~14698 is very 'TESS-like'
compared to the other green points due to its short outer period.

What is probably the most interesting finding from our sample of new stars is the fact that a discovery
of six new apsidal-motion triples can still be done with a modest technique and still on relatively
bright stars. The brightest system in our sample (NSV~14698) has its magnitude below 11 mag, which is
suitable for obtaining good spectra even with a moderate-sized telescope. Moreover, this system shows a
fast apsidal motion together with a short ETV period, and it should be studied in detail due to its
prospective complex dynamical behaviour. Therefore, we call for special attention on this star as well
as for new observations, especially of this interesting target.

\begin{acknowledgements}
We would like to thank an anonymous referee for his/her useful suggestions greatly improving the whole
manuscript. We do thank the NSVS, Kelt, ZTF, ASAS-SN, SWASP, and TESS teams for making all of the
observations easily public available. The research of PZ and MW was partially supported by the project
{\sc Cooperatio - Physics} of the Charles University in Prague. This work has made use of data from the
European Space Agency (ESA) mission {\it Gaia} (\url{https://www.cosmos.esa.int/gaia}), processed by
the {\it Gaia} Data Processing and Analysis Consortium (DPAC,
\url{https://www.cosmos.esa.int/web/gaia/dpac/consortium}). Funding for the DPAC has been provided by
national institutions, in particular the institutions participating in the {\it Gaia} Multilateral
Agreement. This research made use of Lightkurve, a Python package for TESS data analysis
\citep{2018ascl.soft12013L}. This research has made use of the SIMBAD and VIZIER databases, operated at
CDS, Strasbourg, France and of NASA Astrophysics Data System Bibliographic Services.
\end{acknowledgements}

\end{document}